\documentclass{article}
\usepackage{ijcai24}
\usepackage{times}
\usepackage{soul}
\usepackage{url}
\usepackage{multirow} 
\usepackage{array} 
\usepackage[hidelinks]{hyperref}
\usepackage[utf8]{inputenc}
\usepackage[small]{caption}
\usepackage{graphicx}
\usepackage{amsmath}
\usepackage{amsthm}
\usepackage{listings}
\usepackage{booktabs}
\usepackage{algorithm}
\usepackage{algorithmic}
\usepackage[switch]{lineno}

\title{Leveraging Large Language Models for enhanced personalised user experience in Smart Homes}

\author{
Jordan Rey-Jouanchicot$^{1,2,3}$,
André Bottaro$^3$,
Eric Campo$^2$,
Jean-Léon Bouraoui$^3$,
Nadine Vigouroux$^1$,
Frédéric Vella$^1$\\
\affiliations
$^1$IRIT, University of Toulouse, CNRS, Toulouse INP, UT3, UT2J, Toulouse, France\\
$^2$LAAS-CNRS, University of Toulouse, CNRS, UT2J, Toulouse, France\\
$^3$Orange Innovation, Blagnac, 31700, France\\
}
\begin{document}

\maketitle

\begin{abstract}
Smart home automation systems aim to improve the comfort and convenience of users in their living environment. However, adapting automation to user needs remains a challenge. Indeed, many systems still rely on hand-crafted routines for each smart object.

This paper presents an original smart home architecture leveraging Large Language Models (LLMs) and user preferences to push the boundaries of personalisation and intuitiveness in the home environment.
This article explores a human-centred approach that uses the general knowledge provided by LLMs to learn and facilitate interactions with the environment.

The advantages of the proposed model are demonstrated on a set of scenarios, as well as a comparative analysis with various LLM implementations. Some metrics are assessed to determine the system's ability to maintain comfort, safety, and user preferences. The paper details the approach to real-world implementation and evaluation.

The proposed approach of using preferences shows up to 52.3\% increase in average grade, and with an average processing time reduced by 35.6\% on Starling 7B Alpha LLM. In addition, performance is 26.4\% better than the results of the larger models without preferences, with processing time almost 20 times faster.

\end{abstract}

\section{Introduction}

Networks of devices are deployed to assist human beings in their daily activities, using notions of context and knowledge to decide on the best actions to take \cite{weiser1991ubicomp} \cite{dey_understanding_2001}. Indeed, many houses are covered by wireless and wired networks and equipped with electronic devices allowing the occupants to control their environment for comfort, entertainment, security, energy management, and elderly care. 

However, Smart Home Automation Systems are still missing the aim of autonomously taking the best action in every situation. Aligning automation routines to meet every need for every home configuration, every set of devices, available or not, functional or not, remains a challenge. In fact, most systems today are configured for simple routines that occur frequently. For instance, setting the home for wake-up or departure time by playing music, ringing a bell, acting on lights, shutters, heating, ventilation, and air conditioning. More precise needs in less frequent situations are not covered. Furthermore, if some devices are missing from a routine, no decision is made to use alternative devices. While artificial intelligence can play a role in learning about situations and the associated actions taken by users \cite{rashidi_keeping_2009}, it still requires time to understand users' habits, and is never able to cover the wide range of situations encountered at home.

One of the main technical challenges of pervasive computing is the ability to set up a system knowing the wide variety of users’ potential needs, and how it can adapt to the wide range of functions of existing devices, with devices and locations whose configuration can be very different. As their execution environment is particularly dynamic, applications need to be aware of their context and act appropriately. 

Large language models \cite{radford2019language} have the potential to give this general knowledge at once to applications such as smart home. In essence, these models have acquired a vast amount of knowledge. They are trained on a diverse range of textual sources, enabling them to cover a variety of topics, facts, and concepts, here the expected actions of home devices to meet user needs in a wide range of situations. 

Retrieval-Augmented Generation \cite{rag} is a technique that improves the accuracy of LLMs by giving them access to more targeted and precise information.
This is achieved by using a retrieval component that searches a specific database and introduces it into the model, along with users' query.


This paper proposes a software architecture integrating the general LLM knowledge available today into a smart home automation system. The LLM is placed at the center of the home's decision-making system and participates in the reaction to every event to deduce the next best actions. This paper investigates the inclusion of preferences with a LLM for smart home automation, 
The contribution also includes a user-centred representation of smart home states and actions in natural language. Finally, experiments are carried out using several LLMs with different prompting styles for decision-making in the smart home.

The next section presents related work in the literature. Then, the proposed software architecture is detailed, in particular the integration of LLMs for decision-making combined with RAG for user context retrieval. Afterwards, the experimentation section describes a dedicated benchmark of LLMs and prompting styles with a set of home scenarios, leading to the results in the next section. These results are discussed and a final section is devoted to the conclusion and future work. 

\section{Related works}
This topic is recent and few papers have been published on LLMs for smart homes. The following papers are selected for their approaches using the knowledge of user preferences on decision-making in smart home automation systems. This section cannot be exhaustive in this larger domain. 

\cite{oliveira} proposes a multi-agent environment with a Belief-Desire-Intention cognitive model, to support adaptivity and preferences transparently in a smart home environment. Any possible interaction could be modeled in such a way as to allow alternative proposals, but this requires considerable work in semantic representation to be fully complete.

Another paper takes advantage of general knowledge information to improve system adaptivity to preferences. \cite{Ruiza} proposes an approach using three Knowledge-Based systems, one with general knowledge, one with skill knowledge, and one with contextual information such as device location, and generating rule models for a middleware platform based on all this information.

This approach requires really strict semantic modeling to handle most scenarios, and cannot take advantage of some information that is implicitly given in context.

\cite{shuvo} proposes an actor-to-critic (A2C)-based algorithm adapted to decision-making in smart homes for energy consumption.
In this work, at each step, an A2C algorithm is applied to each device to select the best action, using as inputs the activity and price of electricity at that time.

A key element is that the set of actions for each appliance depends on the category associated with the appliance, adding initial knowledge to the model to ensure action based on the importance of the appliance.
This system requires learning, and any change in device availability leading to different optimal decisions will require a large number of steps before adapting.

In addition, the larger the context, the more difficult it will be to converge for each scenario to learn the best-suited device state.
The paper HESIDR, \cite{zhang} also proposes a system for energy consumption in smart homes, it combines a set of control rules and reinforcement learning to reduce the adaptation time.

The study proposed by \cite{peng} describes an approach for decision-making in home automation using deep reinforcement learning. It showed the ability to learn when to turn on a light but with a really limited context supported, so the application is limited to learning when to turn on a light with schedules of 15 minutes, which is a quite large period for this application.

\cite{king_get_2023,king_sasha_2023} proposes the first approach to smart home automation, using a JSON data representation from a smart home middleware platform and experimenting with an LLM to select an action based on a user request.

This approach is a first step towards the use of the general knowledge provided by these models. However, it does not support user preferences, and the idea is to select actions based on an initial user request, rather than proactively. These works have proposed ways of managing decision-making, but are limited by the contextual data supported. Works using symbolic AI methods can show great adaptability, but at the cost of extensive semantic modeling and with requirements to make them adaptable to future changes in preferences. A new method is proposed to support contextual data while adding preferences.

\section{Proposed architecture}
This study proposes a new architecture for decision-making in smart home automation systems.
The system uses Large Language Models and proposes methods to add user preferences, in order to select an action according to context and users. It aims to be a proactive system. At every event occurring in the home, the system proposes actions on devices to align the home state with user needs and preferences.
The system supports different types of data thanks to LLMs ability to process data while generating a textual representation of the home based on device states and the action list.

The proposed system aims to be able to adapt to changes in user preferences over time. Indeed, LLMs avoid the need for retraining and handle appliance configuration changes, basic unavailability, or appliance failure.

This reason leads directly to the use of RAG or directly injecting the knowledge into the prompts for retrieving updated preferences at every execution time of the AI, instead of fine-tuning. Indeed, fine-tuning over preferences would be impractical, as it would require a new iteration each time user preferences change, which would be costly and computationally expensive.

A home simulator is implemented, it takes information on the configurations of sensors and then generates a textual representation, it is also used to generate the list of actions from its data.

\begin{figure}[h!]
    \centering
    \includegraphics[width=8cm]{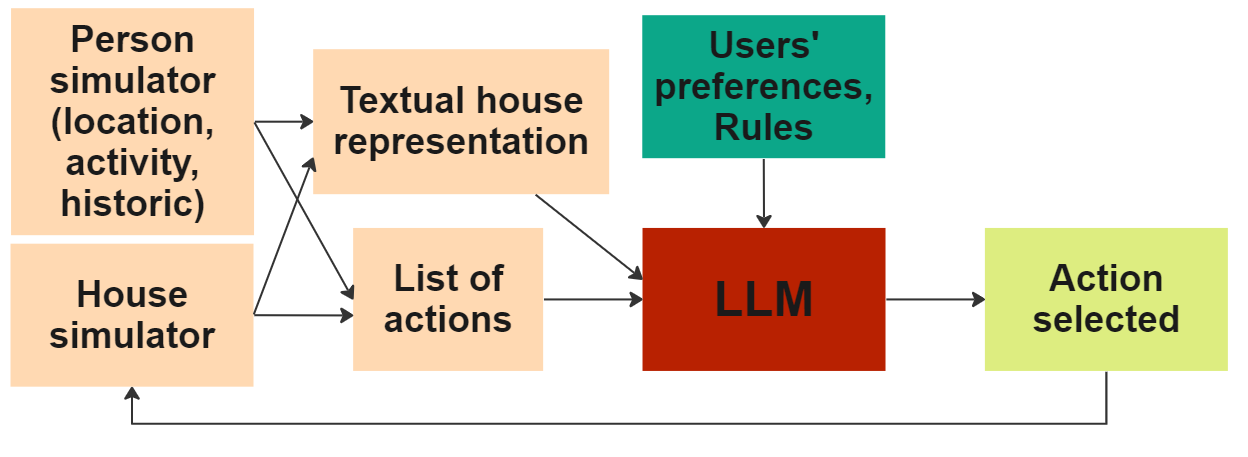}
    \caption{The implemented architecture}
    \label{fig:archi}
\end{figure}

This section details the main components designed for the proposed system architecture. Figure \ref{fig:archi} shows this architecture, including the simulator.\\

The system generates a user-centred text description of the home and a list of actions, with control over connected devices. It filters the number of relevant actions that can be taken in every situation. For example, it limits some actions that may be prohibited to guarantee user safety. This list of actions is used by the model to select the optimal action.

Concerning contextual data representation editing:
\begin{itemize}
    \item User positions are listed with their current activity and the history of previous activities.
    \item The history of previous actions performed in the house is supported.
    \item All rooms, sensors, and actuators are presented using their names, which the user optionally gives.
    \item Some sensors and actuators give more global data and control, e.g., internal and external temperature sensors, gas sensors, humidity sensors, HVAC systems, etc.
\end{itemize}

The implementation supports some device categories with a dedicated natural representation, to generate more natural sentences adapted to some types of data: lights, CO2 sensors, smart curtains, etc. 
It could also support any additional data sensors with a generic representation template, using the device name in the smart home environment and data status.
Using meaningful naming added by the users helps the system to understand the usage of the device.

The action proposal algorithm \ref{alg:listbuilder} considerably reduces the set of possible actions. It assumes that only devices that are in the room or global can be switched on, but that all switched-on devices can be switched off.
As far as the list of possible actions is concerned, the idea is to filter the actions supported according to some conditions and device types. This approach is made possible by LLM's native support for a change in the output action space, without requiring training.

\begin{algorithm}
\begin{algorithmic}[1]
\caption{Dynamic Devices: Action builder Algorithm}
\label{alg:listbuilder}
\REQUIRE $userid$, $devices\_list$
\FORALL{$device\_name$, $device\_kind$, $device\_location$, $device\_state$ in $devices\_list$}
\IF{$device\_kind = "actuator"$}
\IF{$user\_location[userid]$ = $device\_location$ or $device\_state$ = $1$}
\STATE $devices.append(device\_name$ \textit{is} $device\_state)$
\STATE $action\_vector.append(1)$
\ENDIF
\ENDIF
\ENDFOR
\STATE $devices.append(\text{"Interact with user"})$
\STATE $action\_vector.append(2)$
\STATE $devices.append(\text{"No action required"})$
\STATE $action\_vector.append(0)$
\RETURN $(action\_vector, devices)$
\end{algorithmic}
\end{algorithm}

Any type of device can be easily supported: it simply has to be added to the representation and the LLM will ingest the data thanks to its internal knowledge.
This knowledge allows the LLM to get a natural human description of the home including biased contextual data unlike many conventional home automation systems. The latter do not take advantages of information such as the names of lights or rooms. 
This data is transmitted to the LLM using one of the prompting styles,
prompting being the way to call the LLM with contextual arguments. The different styles of prompts will be described in the following section.

A common aim for all the prompting styles is to take advantage of the knowledge of the overall world provided by this Large Language model, to handle changes or even new types of sensors added to the representation.

The user preferences and rules block represents a database containing information about the system's basic rules, generalities about human preferences and specific user preferences.

A benchmark is established with predefined scenarios for evaluation purposes.

Regarding LLMs, some off-the-shelf models are selected, with a local inference engine backend.

\section{Experimentation}
Different objectives are defined for the experimentations.
\begin{itemize}
    \item Evaluate the improvements provided by adding user preferences, with various techniques of doing so.
    \item Evaluate the improvements of natural language representation of a smart home automation state over a JSON representation, as LLMs are trained on natural language corpus. Even if they contain other kinds of data like code.
\end{itemize}

In alignment with the objectives of experimentations, different metrics are used for these evaluations:
\begin{itemize}
    \item Grades: The grade obtained for each model, the prompting styles for each scenario, by execution.
    , based on the grade values defined in table \ref{table:scenarios_metrics}
    \item Processing time: Total runtime, including the construction of the context data and action list representation, the inferences with the prompting styles, the use of RAG if the prompting style uses it, and the processing of the formatted LLM response.
\end{itemize}

Eleven evaluation scenarios are defined as starting points, with predefined accepted actions and their specific grade.
Table \ref{table:scenarios_metrics} presents the eleven scenarios by name and the associated reward values.
A category is associated with each scenario, the goal being to regroup the scenario with the name of the main evaluated ability.
\begin{table}[h!]
\scriptsize
\centering
\begin{tabular}{|p{1.2cm}|p{0.5cm}|p{4.2cm}|p{0.9cm}|}
\hline
\textbf{Scenario Name} & \textbf{Grade} & \textbf{Associated Answer} & \textbf{Category} \\
\hline
\multirow{3}{=}{Out of bed at night} & 2 & Turn on auxiliary light or main light with reduced luminosity level & Safety \\
& 1 & Turn on main light & \\
& 0 & Everything else & \\
\hline
\multirow{3}{=}{Watching TV: late evening} & 2 & Turn on auxiliary light or main light with reduced luminosity level & Comfort \\
 & 1 & Turn on main light, open curtains, discuss & \\
 & 0 & Everything else & \\
\hline
\multirow{3}{=}{Out from bed issue with CO2} & 2 & Inform user of risk & Safety \\
 & 1 & Do an action and inform the user of risk & \\
 & 0 & Everything else &\\
\hline
\multirow{3}{=}{Going back to bed at night} & 2 & Turn on auxiliary light or main light with reduced luminosity level & Safety \\
& 1 & Turn on main light & \\
& 0 & Everything else & \\
\hline
\multirow{3}{=}{Evening sleeping: TV ON} & 2 & Turn off TV & Preference \\
 & 1 & Turn off anything on & \\
 & 0 & Everything else & \\
\hline
\multirow{3}{=}{At dinner watching TV} & 2 & Turn on auxiliary light or main light with reduced luminosity level, open curtains & Preference \\
 & 1 & Turn off the main light, do nothing & \\
 & 0 & Everything else & \\
\hline
\multirow[t]{3}{=}{Forgot to turn off TV: user out} & 2 & Turn off TV, turn off HVAC & Comfort \\
 & 1 & Turn off all lights & \\
 & 0 & Everything else & \\
\hline
\multirow[t]{3}{=}{Too low temperature} & 2 & Turn on HVAC & Preference \\
 & 1 & Open Curtains & \\
 & 0 & Everything else & \\
\hline
\multirow{3}{=}{Low luminosity day} & 2 & Open curtains & Preference \\
 & 1 & Turn on any light in the room & \\
 & 0 & Everything else & \\
\hline
\multirow{3}{=}{Failed curtains} & 2 & Turn on any light of the room & Comfort \\
 & 1 & Open curtains & \\
 & 0 & Everything else & \\
\hline
\multirow{2}{=}{Forgot to turn off lights} & 2 & Turn off any lights, all lights, or HVAC & Preference \\
& 1 & &\\
 & 0 & Everything else & \\
\hline
\end{tabular}
\caption{Scenario responses with grades, associated answers, and evaluation categories.}
\label{table:scenarios_metrics}
\end{table}

Figure \ref{fig:context} shows an example on scenario 1 of the generated contextual representation (textual) transmitted to the LLM.\\

The database of preferences and rules is defined in a single file for all scenarios. These data are naturally written sentences, and at the end of each one, information about the style is recorded:  Rules, Preferences, Generality.
The idea is to transmit to the LLM the importance of each data through keywords.
Generality is considered the least important, Preferences the second most important, and Rules the most important.
The database includes some preferences, generality and some rules. It is designed to handle some scenarios, help in some others but does not provide a solution for all scenarios.
The data are fed into a vector database so that RAG may be used instead of prompts to convey them to the LLM.

Four different prompting styles are compared on all the scenarios:
\begin{itemize}
    \item direct: A system prompt and a prompt to request direct answers in the specified format.
    \item directPref: A system prompt with the preferences, rules and generality from the database and a prompt to request answers in the specified format.
    \item OpenQuestion: Two-steps chain: A system prompt, and a prompt to request: "a list of 3 main problems". For each of the 3 problems uses RAG to get the 3 closest preferences. A prompt to request answers in the specified format
    \item ThreeQuestion: Three-steps chain: A system prompt,a prompt to request: "a list of 3 main problems". For each of the 3 problems uses RAG to get the 3 closest preferences. A prompt to prompt to request answers in the specified format requests (2 times). A final prompt to request an answer in the specified format (based on the previous ones).
\end{itemize}

The database and prompting styles are available in the supplementary material.

A common point between all the prompting styles is the action expected in the output: as mentioned above, most supported devices, such as lights or HVAC systems, are considered as switches in the action list; therefore, in the output of all prompting styles in addition to a "reasoning" and an "action" key, three optional keys are available: temperature, luminosity and explanation. It enables the model to respectively modify the temperature of an HVAC system when executing a related action, modify the luminosity by dimming a light or give an explanation to transmit a sentence to the user.

Two ways of representing the state of the house data are implemented, both using the same input data:
\begin{itemize}
    \item JSON: A JSON representation 
    \item Textual: A fully natural textual representation
\end{itemize}

The implementation of the system is evaluated using various open-sources LLMs, including:
\begin{itemize}
    \item Starling Alpha 7B \cite{starling2023}- 8bpw
    \item Qwen 1.5 14B \cite{qwen} - 5bpw
    \item Qwen 1.5 72B \cite{qwen} - 3.5bpw
\end{itemize} 
The three models are selected for their performance and for covering the three main open-source model sizes. They are used to evaluate the impact of proposed prompting methods and data representation. 

Qwen 1.5 72B, with around 72 billion parameters, is currently one of the best models available open-source. Starling 7B Alpha, with around 7 billion parameters, is an excellent smaller model, and is based on Mistral 7B an efficient model for its size on various benchmarks. Qwen 1.5 14B model, a smaller version of Qwen 1.5 72B, is selected to add an intermediary model.\\

All these models are used with versions that are quantized\footnote{\label{n1} \url{https://github.com/turboderp/exllamav2}}, a technique used to reduce inference time and memory footprint, the quantization chosen for each model is given in bits per weight (bpw).
With their quantization, they require around 8GB, 12GB, and 44GB of memory respectively.

Every model is evaluated on local instances, served locally with an engine-based API backend, using TabbyAPI\footnote{\url{https://github.com/theroyallab/tabbyAPI}} based on ExLLamaV2\footref{n1} multiple GPUs and without automatic splitting, using a workstation equipped with a Ryzen 9 7950x, 96GB of DDR5 memory running at 5600mhz and 2 Nvidia RTX 4090, each with 24 GB dedicated memory, running Ubuntu 23.10.

Experiments are carried out beforehand on various uncontrolled scenarios to define LLM parameters. With the sole aim of reducing non-determinism from one cycle to the next, the final parameters modified from the default engine parameters are as follows:
\begin{itemize}
    \item max\_tokens, maximum number of tokens in output: 300 
    \item min\_p, minimum percentage value that a token must reach to be considered (Value is scaled based maximum token probability): 0.05
    \item temperature, parameter that regulates the randomness: 0.2
\end{itemize}

The RAG is implemented using Langchain\cite{langchain} with an inference engine from HuggingFace\footnote{\url{https://github.com/huggingface/text-embeddings-inference}}, to locally execute an embedding model:
BAAI/bge-large-en-v1.5\cite{bge_embedding}, and an Elasticsearch\footnote{\url{https://github.com/elastic/elasticsearch}} local instance is used as vector database, both instances running on CPU and associated memory.\\

To evaluate system performance, each scenario is executed 10 times with each prompting style, and a grade is given to each response. 
Results then count the total number of points for each prompting category in general, and also for each defined metric associated with each question. The system's complete processing time is also measured, in order to estimate the average latency of the different prompting styles.

The theoretical random action grade for each scenario is calculated as a baseline, using the following formula \ref{eq:grade}.\\
\begin{equation}
\scriptsize
\text{grade}_s = \frac{\text{number\_of\_actions\_rated\_1}_s + 2 \times \text{number\_of\_actions\_rated\_2}_s}{\text{number\_of\_actions}_s}
\label{eq:grade}
\end{equation}
Figure \ref{fig:context} shows an example of a scenario data representation using proposed natural language textual representation.
\begin{figure}[h!]
\scriptsize
\begin{lstlisting}[mathescape]
Current State of the House:
User 1 is in the Livingroom.
User is watching TV.
Previously: User was currently looking at TV

Livingroom: Curtains are Closed. 
Lights: main, floor lamp are respectively Off, Off.
There is a TV in the room and its state is on. 
CO2 level in room is 513ppm.
.......................................
Kitchen: Curtains are Closed. 
Lights: main are Off. 
CO2 level in room is 473ppm.

House was cleaned today. 
Expected cleaning one time a week.
Centralized HVAC system is on with objective to 20${^\circ}$C.
Entrance smart Door is locked.
Time: 10:21 PM
Global house temperature is 20${^\circ}$C,
outside temperature is 5${^\circ}$C.
\end{lstlisting}
\caption{Extract of natural data representation on the scenario "Out of bed at night"}
\label{fig:context}
\end{figure}

As previously mentioned, several data types have been included in the representations: numerical values, boolean values, and character strings. Using an LLM to process context allows the system to support all these data types.\\

Regarding the action list building algorithm, in one of the scenarios for example, it reduces the set of actions from more than 18 to 6 actions, which represents a 3-fold reduced set in this case. This reduced set takes into account the fact that lighting appliances are basic switches and cannot be set with different luminosity levels.

This limited set of actions makes it interesting to use a model randomly selecting one action from the set as a baseline. With a reduced action space, a random answer may not be so bad.
\section{Results}
\subsection{Data representation}

The first analysis of the results focuses on the advantage of using a natural representation versus a JSON representation.
\begin{table}[h!]
    \centering
    \begin{tabular}{c|c|c}
         \bf{Model} & \bf{JSON} & \bf{Natural} \\
         Qwen 1.5 72B & 1.25 & 1.18 \\
         Qwen 1.5 14B & 0.63 & 0.99 \\
         Starling Alpha 7B & 1.03 & 1.18 \\
    \end{tabular}
    \caption{Comparing the two contextual representations: average grade of models per execution per scenario}
    \label{tab:jsonrep}
\end{table} 

Table \ref{tab:jsonrep} shows an average difference between results using JSON representation, or the more natural representation. It shows that the larger models are almost stable independently of the contextual representations, and on average even slightly better using JSON representation, by around 5.9\%. However, on smaller models, the natural language representation greatly improves performances, with a 14.6\% increase in average performance using Starling 7B, and a 57.1\% increase using Qwen 1.5 14B. In terms of processing time, results are quite similar on average for models with both representations. On average, accross all models, it leads to an increase in performance of 21.9\%, despite the results on the larger Qwen model, making the natural representation more efficient.

\subsection{User preferences}
\begin{figure*}
    \centering
    \includegraphics[width=15cm]{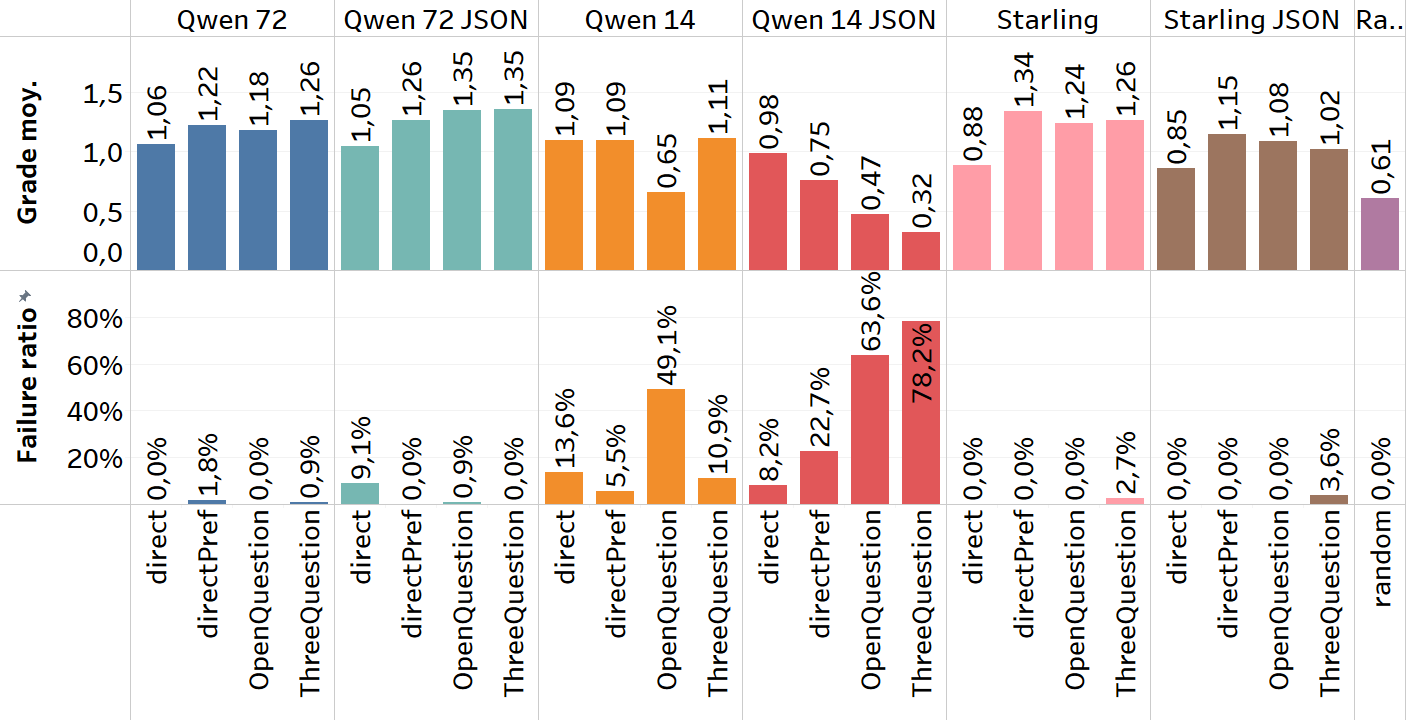}
    \caption{Average grades by model, data representation and prompting styles}
    \label{fig:grade}
\end{figure*}
Figure \ref{fig:grade} depicts the average grades obtained by the 3 chosen models on their responses to scenarios with the 4 different prompting styles and the 2 distinct representation types.

It first shows that Qwen 72B model is much more stable than the other two and that the prompting style does not have as much impact on response quality. Regarding the two smaller models, larger inconsistency in results can be noted as varying with the chosen prompting styles, particularly with Qwen 14B model.

The results of LLMs are compared with a baseline corresponding to the random choice of an action. The results remain on average behind any model without preferences (37.1\% below the worst result with the "direct" prompting style). However, thanks to the algorithm reducing the set of actions and the fact that multiple responses are acceptable on each scenario, the random baseline obtains grades that are sometimes better than the ones of some experiments with LLMs.

The sequence of multiple questions requires the model to be consistent and to respect the expected instruction format. Furthermore, given that only prompt engineering is used to ensure the format, some prompting styles with multiple questions may lead to invalid responses, forcing the model to take a default action in the proposed setting. This default action is set to do nothing and to inform the user that it has failed to act. This reduces the performance of some models with some prompting styles. in figure \ref{fig:grade} failure ratio measures the ratio of invalid responses. 
Qwen 14B results are especially below this baseline because they failed to answer in a large number of scenarios

On average, "directPref" prompt models achieved a gain of 11.3\% over basic prompt with natural representation, and even 20.0\% with JSON representation.

The best results are achieved with JSON representation for "OpenQuestion" and "ThreeQuestion", leading to a 28.6\% improvement over the "direct" prompting style.

With the natural representation, Qwen 14B model gives almost stable results with all prompting styles except "OpenQuestion", and the main difference with "OpenQuestion" seems to be linked to a high failure rate, as shown in figure \ref{fig:grade}.
With JSON representation, failures are so high with all advanced prompting styles that the best results are obtained with the most basic prompt. In this case, "direct" prompting style is 30.7\% more efficient than the "directPref" prompting style.

On Starling 7B, with both representations, the best results are obtained with "directPref", on the Starling representation, its results are 52.3\% better than the direct representation and respectively 8.1\% and 6.3\% better than "OpenQuestion" and "ThreeQuestion" prompting styles.

In addition, the average processing time is reduced by 35.6\% on Starling 7B Alpha using the "directPref" prompting style instead of "direct".

With JSON representation, the averaged grade results follow the same trend, but with lower overall values.\\

Table \ref{fig:grade} shows that RAG-based prompting models can lead to better results than a single prompt containing all the data, as it is visible Qwen 72B.
However, to date, LLM inference is still slow, and the use of complex prompts leads to a loss of accuracy in the test scenarios with smaller models.\\

\begin{table}[h!]
\scriptsize
\centering
\begin{tabular}{|l|l|l|l|}
\hline
Model & \textbf{Prompting Style} & \multicolumn{1}{c|}{\textbf{Average grade}} & \multicolumn{1}{c|}{\textbf{Proces. time (s)}} \\ \hline
\multirow{4}{*}{\textbf{Qwen 72}}       & direct        & 1,06 & 9,04  \\ \cline{2-4} 
                                        & directPref    & 1,22 & 7,01  \\ \cline{2-4} 
                                        & OpenQuestion  & 1,18 & 24,43 \\ \cline{2-4} 
                                        & ThreeQuestion & 1,26 & 42,13 \\ \hline
\multirow{4}{*}{\textbf{Qwen 72 JSON}}  & direct        & 1,05 & 9,02  \\ \cline{2-4} 
                                        & directPref    & 1,26 & 6,71  \\ \cline{2-4} 
                                        & OpenQuestion  & 1,35 & 25,20 \\ \cline{2-4} 
                                        & ThreeQuestion & 1,35 & 41,47 \\ \hline
\multirow{4}{*}{\textbf{Qwen 14}}       & direct        & 1,09 & 1,22  \\ \cline{2-4} 
                                        & directPref    & 1,09 & 1,16  \\ \cline{2-4} 
                                        & OpenQuestion  & 0,65 & 8,83  \\ \cline{2-4} 
                                        & ThreeQuestion & 1,11 & 15,54 \\ \hline
\multirow{4}{*}{\textbf{Qwen 14 JSON}}  & direct        & 0,98 & 1,28  \\ \cline{2-4} 
                                        & directPref    & 0,75 & 1,21  \\ \cline{2-4} 
                                        & OpenQuestion  & 0,47 & 9,39  \\ \cline{2-4} 
                                        & ThreeQuestion & 0,32 & 16,62 \\ \hline
\multirow{4}{*}{\textbf{Starling}}      & direct        & 0,88 & 0,73  \\ \cline{2-4} 
                                        & directPref    & 1,34 & 0,47  \\ \cline{2-4} 
                                        & OpenQuestion  & 1,24 & 3,98  \\ \cline{2-4} 
                                        & ThreeQuestion & 1,26 & 6,23  \\ \hline
\multirow{4}{*}{\textbf{Starling JSON}} & direct        & 0,85 & 0,48  \\ \cline{2-4} 
                                        & directPref    & 1,15 & 0,48  \\ \cline{2-4} 
                                        & OpenQuestion  & 1,08 & 3,77  \\ \cline{2-4} 
                                        & ThreeQuestion & 1,02 & 6,20  \\ \hline
\end{tabular}
\caption{Comparison of average grades for each prompting style and models with inference time}
\label{tab:inf}
\end{table}

Adding inference time to the balance with table \ref{tab:inf} highlights Starling 7B Alpha results with "directPref". It outperforms almost all others except Qwen 72B with JSON representation and prompt chaining, but is 53.6 times faster, with a lower grade of just 0.7\%.

With Qwen 72B, the use of more complex models (OpenQuestion, ThreeQuestion) can lead to better results, as seen previously in particular with JSON representation, but this comes with a trade-off: inference time. 
Qwen 72B model is already much slower due to its number of parameters, and due to the number of operations required to produce a single token. For instance, an inference with "directPref" takes 6.86 seconds on average (JSON and textual representation), whereas using "OpenQuestion" (which is around 60\% faster than "ThreeQuestion") is 3.6 times slower.

In the current state of this type of hardware, it is impossible to consider them as a viable alternative for managing a smart home automation system, with such reaction times.

\section{Discussion}
This section mainly discusses the results of the LLMs and prompting techniques chosen in order to make choices for real experiments. The advantages of this study are highlighted as well as the new challenges that are raised.\\

As seen previously, using larger LLM such as Qwen 72B allows greater stability in preference-free scenarios. Indeed, Qwen 72 is 19.9\% better than Starling 7B in this case. However there are drawbacks, the first being the inference time as mentioned, and the second being the hardware infrastructure required. The quantized version of Qwen 1.5 72B requires 44GB of memory compared with around 8GB for Starling 7B Alpha with lower quantization.

Compared with Qwen 72B model using JSON representation, Starling 7B Alpha with "directPref" takes advantage of natural representation and achieves almost similar performances with much more complex prompting techniques.
This makes the approach of using Starling 7B Alpha with this prompting style a good choice for future work. It gives similar performances concerning grades, and has a relatively low inference time (Average: 0.47s).

In addition, compared to Qwen 1.5 72B with natural representation and no preferences, Starling 7B Alpha's performance is 26.4\% better using "directPref", with a processing time almost 20 times faster.\\

The results show the advantage of adhering to preferences. Drawbacks appear, however, with the additional average computation time for "OpenQuestion" and "ThreeQuestion" prompting styles, which use RAG.
Using RAG brings no advantage in most cases. This is certainly due to the relatively small database. If the system required a larger database of preferences and rules, the results might have been different as it would not have been possible to give them directly through "directPref" prompting style. Based on current results, the best choice for a use case with a larger database would remain Starling 7B Alpha, with "OpenQuestion" prompting style and natural representation, as it provides results that are aligned with users preferences, with only 8.1\% less average grade than Qwen 1.5 72B, while keeping an acceptable average inference time (3.98 seconds vs 25.20 seconds).

The use of smaller models poses the challenge of enforcing the output format. New methods have recently been proposed such as Outlines\footnote{\url{https://github.com/outlines-dev/outlines}}, to strengthen the output grammar. They should be tested soon in order to analyze the cost in inference time and the improvement of the results of this paper.

Another approach is fine-tuning. The limits of using fine-tuning for user preferences were mentioned previously. However, fine-tuning on smart home domain data could allow the system to ensure the output format while improving the analysis of contextual data.\\

Although the proposed system is not comparable to other works due to its completely new approach, it has the advantages of its nature, as well as drawbacks. Firstly, it is not deterministic. Each configuration has been run 10 times, but some results are different each time, which means that the results of the system cannot be certified and a safety layer must be developed to secure some actions. Secondly, LLM queries are slow and require significant computating and memory resources: adaptability without training comes at a cost. 

\section{Conclusion}
This paper presents a new architecture for a smart home automation system, using LLMs with user preferences to enhance personalised user experiences.
This approach leverages the general knowledge provided by LLMs and combines it with naturally written rules and preferences to make contextually relevant decisions in line with user preferences. This architecture is proactive, able to adapt to any change in the environment thanks to the robustness provided by LLMs.

The user-centred action list builder takes advantage of this ability to reduce the set of actions at each step, as the representation of the environment also takes advantage of the name given to devices by the user to better support them.

The experimental results demonstrate the potential of this architecture to improve alignment with user preferences compared with an implementation without user preferences, showing up to 52.2\% performance increase.

The study showed that, particularly with small models, using a natural representation instead of a JSON representation leads to an increase in performance, with an average 21.9\% increase.

Although the system shows promising results on a set of defined scenarios, it also presents challenges due to stochastic behaviour and a slower inference time compared to traditional machine learning methods.
These drawbacks are offset by the system's ability to adapt dynamically - without retraining - to changes in preferences, appliances and home configuration.

Future work will focus on implementing the system in a real-world smart home middleware system such as OpenHAB\cite{openhab} to evaluate its performance with real users. Mechanisms will be proposed to allow users to naturally add and remove preferences, as well as to explore the automatic evolution of these rules and preferences.

\newpage

\appendix

\section*{Ethical Statement}
There are no ethical issues.


\bibliographystyle{named}
\bibliography{ijcai24}

\end{document}